\documentclass[prl,twocolumn,aps,showpacs,superscriptaddress]{revtex4}
\usepackage{epsfig}
\usepackage{graphicx}
\usepackage{color}
\usepackage{amssymb}
\usepackage{amsmath}
\usepackage{bbm}

\usepackage{changes}

\begin{document}

\title{Screening of pair fluctuations in superconductors with coupled shallow
  and deep bands: a route to higher temperature superconductivity}

\author{L.~Salasnich}
\affiliation{Dipartimento di Fisica e Astronomia ``Galileo Galilei",
Universit\`a di Padova, Via Marzolo 8, 35131 Padova, Italy}
\affiliation{Istituto Nazionale di Ottica (INO)
del Consiglio Nazionale delle Ricerche (CNR), Sezione di Sesto
Fiorentino, Via Nello Carrara 2, 50019 Sesto Fiorentino, Italy}

\author{A.~A.~Shanenko}
\affiliation{Departamento de F\'isica, Universidade Federal de
Pernambuco, Av. Jorn. An\'ibal Fernandes, s/n, Cidade
Universit\'aria 50740-560, Recife, PE, Brazil}

\author{A. Vagov}
\affiliation{Institut f\"{u}r Theoretische Physik III, Bayreuth
Universit\"{a}t, Bayreuth 95440, Germany}
\affiliation{ITMO University, 49 Kronverksky Pr., St. Petersburg,
197101, Russia}

\author{J. Albino Aguiar}
\affiliation{Departamento de F\'isica, Universidade Federal de
Pernambuco, Av. Jorn. An\'ibal Fernandes, s/n, Cidade
Universit\'aria 50740-560, Recife, PE, Brazil}

\author{A.~Perali}
\affiliation{School of Pharmacy, Physics Unit, Universit\`{a} di
Camerino, I-62032 - Camerino, Italy}

\begin{abstract}
A combination of strong Cooper pairing and weak superconducting fluctuations is crucial to achieve and stabilize high-$T_c$ superconductivity. We demonstrate that a coexistence of a shallow carrier band with strong pairing and a deep band with weak pairing, together with the Josephson-like pair transfer between the bands to couple the two condensates, realizes an optimal multicomponent superconductivity regime: it preserves strong pairing to generate large gaps and a very high critical temperature but screens the detrimental superconducting fluctuations, thereby suppressing the pseudogap state. Surprisingly, we find that the screening is very efficient even when the inter-band coupling is very small. Thus, a multi-band superconductor with a coherent mixture of condensates in the BCS regime (deep band) and in the BCS-BEC crossover regime (shallow band) offers a promising route to higher critical temperatures.
\end{abstract}


\maketitle

Multi-band and multi-gap superconductors have demonstrated a potential to exhibit novel coherent quantum phenomena that can enhance the pairing energy and the critical temperature $T_c$ ~\cite{mp}. The well-known examples are magnesium diboride~\cite{MgB2a,MgB2b,MgB2c} and iron-based superconductors~\cite{ironA, ironB}, where multiple Fermi surfaces can be effectively controlled by doping or by applying pressure~\cite{bian, okaz}. Multi-band superconductivity can be also achieved in artificial inhomogeneous structures made of a single-band superconducting material - nanofilms, nanostripes or samples with spatially controlled impurity distributions~\cite{str1,str2,str3,str4}.

The phenomenon entangled with the multi-band superconductivity, important in this work, is the BCS-BEC crossover~\cite{eagl,leg,chen,strinati}. Proximity to this crossover in multi-band materials with deep and shallow bands can give rise to a notable increase of superconducting gaps~\cite{shan1, shan2, Innocenti2010, Mazziotti2017}, which on the mean-field level leads to higher $T_c$. The physical reason is the depletion of the Fermi motion in a shallow band, which yields short-sized pairs. In such materials the superconducting state is a coherent mixture of a BCS condensate in deep bands and BCS-BEC-crossover or even nearly BEC condensate in shallow bands. This takes place in, e.g., in ${\rm MgB}_2$~\cite{Innocenti2010}, many of iron-based superconductors~\cite{okaz, guid, bor,kan,kas} and in nanoscale samples~\cite{shan1,shan2}. 

However, the largest enemy of the high-$T_c$ superconductivity in such materials is superconducting fluctuations. They are significant for the same reason which leads to a higher $T_c$ - the depletion of the carrier motion in a shallow band that is associated with a low superconducting stiffness. The fluctuations give rise to the pseudogap state in the interval $T_c <T< T_{c0}$, where the extracted from the mean-field calculations $T_{c0}$ marks the appearance of incoherent and short lived Cooper pairs. The latter develop a coherent state below $T_c$ - the true critical temperature of the superconducting condensate~\cite{sademelo, per}. For shallow bands $T_c\ll T_{c0}$ and this eliminates all gains of the BCS-BEC crossover regime.

In this Letter we consider the mechanism to suppress these fluctuations, which involves the interference of multiple pairing channels with significantly different stiffness. In particular, we investigate how the fluctuation-induced shift in the critical temperature of a superconductor with one shallow and one deep bands is affected by the Josephson-like pair transfer between the bands. Our results demonstrate that even a very small coupling to the stable condensate of a deep band is enough to screen severe superconducting fluctuations and ``kill" the pseudo-gap regime, thereby stabilizing the BCS-BEC-crossover condensate of the shallow band at high temperatures. While a similar effect has been included previously to model the momentum dependent interactions in underdoped cuprates~\cite{per-var} and to study the vortex states~\cite{wolf}, only in the present paper the fluctuation screening is proposed as a key mechanism for stabilizing a higher $T_c$.

Consider the standard microscopic model of a two-band superconductor introduced in~\cite{suhl,mos}, with deep ($\nu =1$) and shallow ($\nu = 2$) bands. The conventional $s$-wave pairing is assumed for both bands, with the symmetric real coupling matrix $\check{g}$ and its elements $g_{\nu \nu^\prime }$. The inter-band coupling $g_{\nu \nu}$ is chosen so that the shallow-band mean-field critical temperature is significantly larger than the deep-band one for the zero Josephson-like coupling $g_{12}=0$. The parabolic dispersion is adopted for both bands $\varepsilon_{\nu,k} = \varepsilon_{\nu,0} + \hbar^2 k^2/2m_\nu$, with $\varepsilon_{\nu,0}$ the lowest energy of the band and $m_\nu$ the band effective mass. For the deep band $\mu \gg \varepsilon_{1,0}$ ($\mu$ is the chemical potential), whereas for the shallow one $\mu \approx\varepsilon_{2,0}$; below we set $\mu = \varepsilon_{2,0}$ for simplicity. For the illustration we choose 2D bands (many multi-band materials exhibit quasi-2D Fermi surfaces). The system is assumed in the clean limit.

The  fluctuations are calculated using the two-component Ginzburg-Landau (GL) free energy functional ~\cite{geilik, kres, ashker, zhit,kogan,shan3,vag,orl,supp}
\begin{align}
& F = \int d^2{\bf r}\Big[\sum_{\nu=1}^2   f_\nu  +  (\vec\Delta, \check{L}
\vec\Delta)\Big],  \notag \\ 
&f_\nu = a_\nu \left|\Delta_{\nu} \right|^2  + \frac{b_{\nu}}{2} \left|
\Delta_{\nu} \right|^4 + {\cal K}_{\nu}  \left| \boldsymbol{\nabla}\Delta_{\nu} 
\right|^2,\notag\\
&\check{L} = \check{g}^{-1} -\left( \begin{array}{cc}  {\cal A}_1 & 0\\
 					0  & {\cal A}_2 \end{array}\right),
\label{eq:GLfreeEN}
\end{align}
where $\vec\Delta = (\Delta_1,\Delta_2)^T$, $(.,.)$ is the scalar product in the band space and the coefficients read as~\cite{wolf}
\begin{align}
&{\cal A}_\nu = N_{\nu} \ln \left( \frac{2 e^{\gamma} \hbar \omega_c}{\pi T_{c0}} 
\right), \,
a_{\nu} = \alpha_\nu (T - T_{c0}), \, \alpha_\nu=  \frac{N_{\nu}}{T_{c0}},  \notag \\ 
&b_{\nu} = \frac{7\zeta(3)}{8\pi^2}\,\frac{N_{\nu}}{T^2_{c0}}, \; 
{\cal K}_{\nu}={\cal M}_{{\cal K},\nu} \, \frac{N_{\nu}\hbar^2v^2_\nu}{T^2_{c0}},
\label{eq:coeff-band}
\end{align}
where $N_\nu$ is the band density of states, $T_{c0}$ is the mean-field critical temperature of the two-band system, $\hbar\omega_c$ denotes the energy cutoff (the same for both bands), $\gamma$ is the Euler constant, $\zeta(x)$ is the Riemann zeta function, ${\cal M}_{{\cal K},1}=\frac{7\zeta(3)}{32\pi^2}, \; {\cal M}_{{\cal K},2} = \frac{3 \zeta(2)}{8 \pi^2}$, and the characteristic band velocities are $v_1 = \sqrt{{2(\mu-\varepsilon_{1,0})/m_1}}$ and $v_2 = \sqrt{2T_{c0}/m_2}$.

$T_{c0}$ is obtained from the linearized gap equation $\check{L}\vec\Delta = 0$~\cite{vag,orl,supp}, solved by $\vec \Delta =\psi \vec\eta$, where $\psi$ is the order parameter and $\vec\eta$ is an eigenvector of $\check{ L}$ with the zero eigenvalue. Then $T_{c0}$ is the largest of the two solutions of $\det\check{L} =0$. The eigenvector $\vec\eta$ can be represented as $\vec\eta = ( S^{-1/2}, S^{1/2})^T $ with~\cite{vag,supp}
\begin{align}
S=\frac{1}{2\lambda_{12}}\!\left[\lambda_{22}-\frac{\lambda_{11}}{\chi} + 
\sqrt{\Big(\lambda_{22}-\frac{\lambda_{11}}{\chi}\Big)^2\!\!+
4\frac{\lambda^2_{12}}{\chi}}\right], 
\label{eq:S}
\end{align}
where $\chi = N_2/N_1$ and $\lambda_{\nu \nu^\prime } = g_{\nu \nu^\prime } (N_1 +  N_2)$. For the critical temperature one obtains
\begin{align}
T_{c0}=\frac{2e^\gamma}{\pi}\hbar\omega_c {\rm exp}\left[-\frac{(1+\chi)
(\lambda_{22}-\lambda_{12}S)}{\lambda_{11}\lambda_{22}-\lambda^2_{12}}\right].
\label{eq:Tc0}
\end{align}

Next we rewrite Eq.~(\ref{eq:GLfreeEN}) using the substitution $\vec\Delta = \psi\vec \eta + \phi\vec\xi$,  where  $\vec\xi $ is orthogonal to $\vec\eta$.  One finds~\cite{supp}
\begin{align}
F  = \int d^2{\bf r}(f_\psi  + f_\phi + f_{\psi \phi}),
\label{eq:GLfreeE}
\end{align}
where 
\begin{align}
 f_\psi  = a_{\psi} \left|\psi  \right|^2  + \frac{b_{\psi}}{2} \left| \psi \right|^4 +
 {\cal K}_{\psi}  \left| \boldsymbol{\nabla}\psi  \right|^2,
\label{eq:f_psi}
\end{align}
while $f_\phi$ is obtained from Eq.~(\ref{eq:f_psi}) by changing $\phi\to \psi$~(also in $a_{\psi}$, $b_{\psi}$, and ${\cal K}_{\psi}$), and $f_{\psi \phi}$ comprises coupling terms between the modes $\psi$ and $\phi$. One notes that only $f_\psi$ describes the critical behaviour of the system while $f_\phi$ adds small non-critical
corrections~\cite{supp}. This difference follows from the fact that $a_\psi= 0$ at $T =T_{c0}$ whereas $a_\phi = (\vec \xi, \check L \vec \xi) \neq 0$ in the same limit~\cite{supp}. Consequently, in the GL domain both the mean field solution~\cite{vag} and the fluctuations are defined by the single-component GL functional $f_\psi$ of Eq.~(\ref{eq:f_psi}). The related coefficients read as~\cite{supp}
\begin{align}
\alpha_{\psi}  = \frac{\alpha_1}{S} + \alpha_2 S, b_{\psi}  = \frac{b_1}{S^2} + b_2
S^2,\, {\cal K}_{\psi}  = \frac{{\cal K}_1}{S} + {\cal K}_2 S ,
\label{eq:coeff}
\end{align}
and $a_{\psi}=\alpha_{\psi} (T - T_{c0})$. 

In weakly coupled superconductors $T_{c0}$ is the superconducting transition temperature at which Cooper pairs are created and form the condensate state. However, in the vicinity of the BCS-BEC crossover the fluctuations destroy the condensate near $T_{c0}$ so that the superconducting transition takes place at $T_c < T_{c0}$. In the interval $T_c < T < T_{c0}$ the system is in the pseudogap regime of incoherent and fluctuating Cooper pairs~\cite{sademelo, per, chen, strinati}.

The actual transition temperature $T_c$ is calculated via the fluctuation correction to $T_{c0}$. To this aim, the order parameter is split into ``slow" (fluctuation averaged part) $\varphi({\bf r})$  and ``fast" (fluctuation) contribution $\eta({\bf r})$  as $\psi({\bf r})=\varphi ({\bf r}) +\eta ({\bf r})$~ \cite{popov}. Then, the fluctuation part of Eq.~(\ref{eq:f_psi}) is approximated by the Gaussian ``Hamiltonian" that is generally written in the
form~\cite{supp} 
\begin{align}
H= \sum\limits_{{\bf q}} \left[ A_{\bf q}(x^2_{\bf q}+ y^2_{\bf q})\! +\! B_{\bf q}(x_{\bf q}x_{-{\bf q}} - y_{\bf q} y_{-{\bf q}})\right],
\label{eq:H}
\end{align}
where $\eta_{\bf q} = x_{\bf q} +  i y_{\bf q}$, with $\eta_{\bf q}$ the Fourier transform of $\eta({\bf r})$, and the coefficients for the diagonal and off-diagonal terms $A_{\bf q}$ and $B_{\bf q}$ depend on a chosen model for the fluctuations. A rough estimate of the fluctuation-driven shift of the critical temperature is given by the Ginzburg number (Ginzburg-Levanyuk parameter) $Gi$ that defines the temperature interval, where the mean field theory is compromised by fluctuations. This estimate neglects the interaction between different fluctuation modes~\cite{lar-var}. We note that a similar approach to the Gross-Pitaevskii equation is called the Bogoliubov approximation~\cite{zaremba}.

Here we use a more involved analysis where the interaction of the fluctuation modes is taken into account~\cite{lar-var} in a way similar to the Popov approximation~\cite{zaremba} for the fluctuation corrections to the Gross-Pitaevskii equation.  In this way, in addition to the linear terms, higher powers of $\eta$ are retained in the GL equation and then linearized within a mean-field approximation. The corresponding fluctuation-averaged GL equation is given by (the subscript $\psi$ of the coefficients $a$, $b$, and ${\cal K}$ is suppressed below)
\begin{align}
\left(a+2b\langle |\eta|^2 \rangle\right)\varphi + b\varphi|\varphi|^2 - {\cal K} \boldsymbol{\nabla}^2\varphi = 0,
\label{eq:GLP}
\end{align}
where $\langle\dots\rangle$ stands for the fluctuation averaging. For the Gaussian ``Hamiltonian" (\ref{eq:H}) we have $\langle\eta\rangle=\langle\eta|\eta|^2 \rangle=\langle\boldsymbol{\nabla}^2\eta\rangle=0$. The anomalous average $\langle\eta^2\rangle$ is zero only when $B_{\bf q}=0$. However, in practical cases $|A_{\bf q}| \gg |B_{\bf q}|$~\cite{lar-var}, which means that $|\langle\eta^2\rangle|\ll \langle |\eta|^2\rangle$ and thus $\langle\eta^2\rangle$ can be ignored (this is equivalent to the random phase approximation).  In our calculations $B_{\bf q}=0$ [see Eq.~(\ref{eq:mf}) below]. 

At $T \geq T_c$ we have $\varphi=0$ and, hence, $\eta$ obeys the standard GL equation resulting from the functional (\ref{eq:f_psi}). To find the corresponding coefficients $A_{\bf q}$ and $B_{\bf q}$, this equation is linearized by invoking the mean-field approximation $\eta|\eta|^2\approx 2\eta\langle|\eta|^2\rangle$, where the factor $2$ ensures the critical enhancement of fluctuations at $T= T_c$. This yields
\begin{align}
A_{\bf q} =a+2b\langle|\eta|^2\rangle + {\cal K}q^2, \;\langle|\eta|^2\rangle = \frac{1}{L^2}\!\sum\limits_{\bf q} \frac{T}{A_{\bf q}},
\label{eq:mf}
\end{align}
while $B_{\bf q}=0$.

The shifted transition temperature is obtained from that the linear term in Eq.~(\ref{eq:GLP}) vanishes, yielding
\begin{align}
T_c = T_{c0} - \frac{2b\langle|\eta|^2\rangle_c}{\alpha},
\label{eq:Tc}
\end{align}
where $\langle |\eta|^2 \rangle_c$ is calculated at $T=T_c$. The fluctuation contribution to Eq.~(\ref{eq:Tc}) is given by a formally divergent integral  
\begin{align}
\langle|\eta|^2\rangle_c=
\int\limits_{\Lambda_0<q<\Lambda_{\infty}}\!\!\!\! \frac{d^2{\bf
q}}{(2\pi)^2} \;\frac{T_c}{{\cal K}q^2} = \frac{T_c}{2\pi{\cal
K}}\ln\left({\Lambda_{\infty}\over\Lambda_0}\right) ,  
\label{eq:etaC}
\end{align}
regularized by the infrared $\Lambda_0$ and ultraviolet $\Lambda_{\infty}$ cut-offs. The ultraviolet cut-off $\Lambda_{\infty} \simeq 1/\xi(0)$ is determined by the applicability of the GL theory, which limits the spatial fluctuation length by the BCS coherence length $\xi(0)$~\cite{lar-var}. The infrared cut-off is related to the upper limit for the phase coherence length that is estimated as the GL coherence length $\xi(T_{Gi})$ calculated at the Ginzburg-Levanyuk temperature $T_{Gi}=T_{c0}(1-Gi)$. Recall that $Gi$ defines the interval in the vicinity of $T_{c0}$, where the mean field theory is compromised by fluctuations and for the case of interest $Gi =b/(4\pi\alpha{\cal K})$. Taking $\Lambda_0=c_0/\xi(Gi)$ and $\Lambda_{\infty} = c_{\infty}/\xi(0)$, with $c_0$ and $c_{\infty}$ constants, and using the standard definition for the GL coherence length $\xi=\sqrt{-{\cal K}/a}$, one obtains \begin{align}
\frac{\delta T_c}{T_c} \equiv\frac{T_{c0} - T_c}{T_c} =  2\,Gi \,\ln\left({1\over 4Gi}\right), 
\label{eq:renormgr}
\end{align}
where we set the ratio $c_{\infty}/c_0=1/2$ in agreement with the renormalization group result as well as with the perturbative approach for the superconducting density~\cite{lar-var}. For $q <\Lambda_{0}$ the system enters the Berezinskii-Kosterlitz-Thouless (BKT) regime of the proliferation of quantized vortices. One can take into account the BKT physics by recalling the Nelson-Kosterlitz criterion~\cite{NKcrit}. In this way the additional shift $(\delta T_c/T_c)_{BKT} \propto Gi$ is obtained~\cite{supp} (see also Fig. 15.1 of \cite{lar-var}). However, in our conditions it produces a very small correction to Eq.~(\ref{eq:renormgr}).

\begin{figure}[t]
\centering
\includegraphics[angle=0,width=0.75\columnwidth]{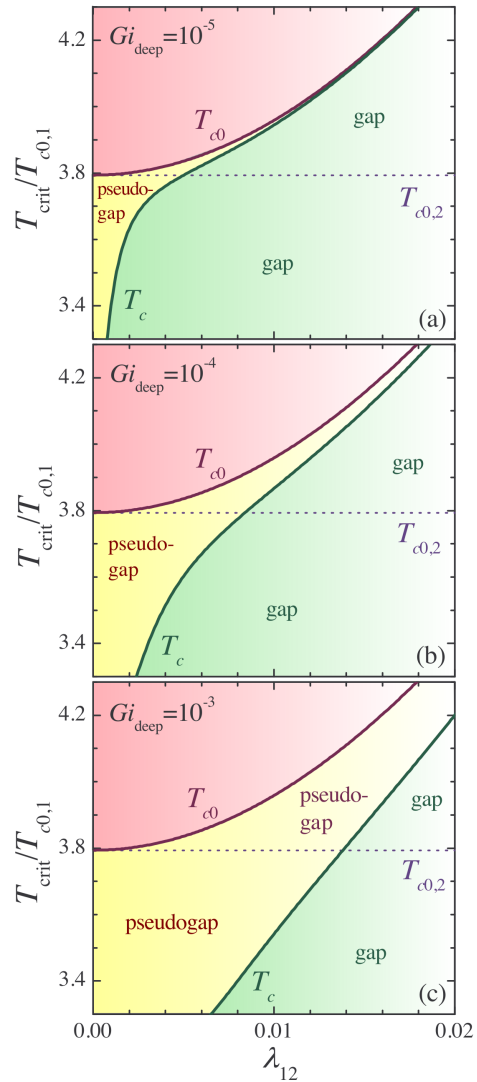}
\caption{The mean-field and fluctuation-shifted critical temperatures $T_{c0}$ and $T_c$ versus the inter-band coupling $\lambda_{12}$. Panels (a)-(c) demonstrate the results for $Gi_{\rm deep} = 10^{-5}$, $10^{-4} $, and $10^{-3}$, respectively. The mean-field transition temperature of the uncoupled shallow band $T_{c0,2}$ is given
as a reference value.} 
\label{fig1}
\end{figure}

In order to investigate the sensitivity of the pair fluctuations to the inter-band coupling, we consider the limit $v_2/v_1 \to 0$~(as $v_2 \ll v_1$) so that $Gi =b/(4\pi\alpha{\cal K})$~[$b$, $\alpha$, and ${\cal K}$ are given by
Eq.~(\ref{eq:coeff}) without subscript $\psi$] is reduced to
\begin{align}
Gi = Gi_{\rm deep} \frac{1 + S^4}{1 + S^2},
\label{eq:Gi}
\end{align}
with $S$ given by Eq.~(\ref{eq:S}) and $Gi_{\rm deep}$ the Ginzburg number of the deep band. The latter depends on $\mu-\varepsilon_{1,0}$ and is a tuneable parameter assuming small enough values. $T_{c0}$ and $T_c$ as functions of $\lambda_{12}$ are found from Eqs.~(\ref{eq:S}),  (\ref{eq:Tc0}),  (\ref{eq:renormgr}), and (\ref{eq:Gi}) , where we use $\lambda_{11}=0.25$ and $\lambda_{22}=0.30$,  and $N_1 = N_2$. Our qualitative conclusions are not sensitive to a particular choice of $\lambda_{11},\,\lambda_{22}$, and $\chi = N_2/N_1$. The only restriction is that the mean-field critical temperature of the uncoupled shallow band $T_{c0,2}$ is significantly larger than that of the deep band $T_{c0,1}$. 

Obtained $T_{c0}$ and $T_c$ are shown in the units of $T_{c0,1}$ in Fig.~\ref{fig1}~($T_{c0,2}$ is also given as a guide for the eye). Three panels are for $Gi_{\rm deep}=10^{-5}$ (a), $10^{-4}$ (b), and $10^{-3}$ (c). The validity range of our results is given by $Gi \lesssim 1/(4e)$, where $1/(4e) \approx 0.092$ is the maximum of the right-hand-side of Eq.~(\ref{eq:renormgr}). At larger $Gi$ our approach cannot be applied, which reflects a huge impact of the fluctuations.

Strikingly, the results in Fig.~\ref{fig1} reveal that the fluctuations are screened almost completely even for extremely small values of $\lambda_{12}$, especially when the fluctuations of the BCS condensate in the deep band are weak enough. For example, for $Gi_{\rm deep} =10^{-5}$~[Fig.~\ref{fig1}(a)] the pseudogap interval becomes negligible at $\lambda_{12} \simeq 0.002 \approx 0.01 \lambda_{22}$. For such small inter-band couplings the two-band superconductor is in the regime of BCS-BEC crossover, governed by the shallow band, but its superconducting critical temperature is close to the mean-field transition temperature of the uncoupled shallow band, i.e., $T_{c0} \simeq T_{c0,2}$. Even for $Gi_{\rm deep} =10^{-4}$ and $10^{-3}$ the pseudogap regime is pronounced only when $\lambda_{12} \ll\lambda_{11},\lambda_{22}$~[Figs.~\ref{fig1}(b) and (c)]. 

Notice, that for real materials $\lambda_{12}$ is usually in the range $0.005 \lesssim \lambda_{12} \lesssim 0.4$ (see Ref.~\cite{vag1} and references therein). Also, the chosen values of $Gi_{\rm deep}$ are in line with conservative estimations for materials with 2D bands~\cite{kett}. Finally, we recall that although our calculations are performed for the case $\mu = \varepsilon_{2,0}$~(for simplicity), increasing the Lifshitz parameter $(\mu-\varepsilon_{2,0})/\varepsilon_{2,0}$ does not change the results even quantitatively as long as $v_1 \gg v_2$ holds.  

In conclusion, this work investigates the pairing fluctuations in a two-band superconductor with one deep and one shallow bands and demonstrates a very effective mechanism to suppress these fluctuations. The shallow band is in the regime of the BCS-BEC crossover, where the superconducting critical temperature is expected to be high. Although superconducting fluctuations in the shallow band alone are very large and would normally destroy the superconductivity, they are screened by the pair transfer inter-band coupling to the deep band. Remarkably, the suppression is very effective, almost complete, even when the inter-band coupling is so small that the superconducting temperature is fully determined by the shallow band. Our results provide a solid explanation of the recent striking observation that the pseudogap was not detected in the multi-band BCS-BEC-crossover superconductor ${\rm FeSe}$, which was called by the authors ``a unique feature that is absent in a single band system"~\cite{Hanag}. Notice that superconductivity is the most robust phase with respect to other instabilities that can arise when shallow bands and high density of states are present in the system.

Our findings open new perspectives in searching for novel multi-band superconductors with higher critical temperatures as the fluctuation screening arising from the interference of multiple pairing channels is a fundamental mechanism to protect the superconductivity. One notes that many other scenarios have been proposed to achieve a high $T_c$ - an interplay between the finite-range pairing potential and the interparticle distance is one of the latest~\cite{Balatsky}. However, all those are based on the mean-field analysis and neglect enhanced fluctuations.
The present mechanism is unique one to avoid severe superconducting fluctuations being an ultimate obstacle toward very high superconducting temperatures.

Finally, we point to the universality of this screening mechanism that can also suppress particle-hole fluctuations in multi-band charge or spin ordered systems, expanding the applicability of our results. For example, it can be relevant for multi-band superfluidity of ultracold fermions in optical lattices~\cite{iskin,tempere} and in multi-orbital fermions with pairing of different channels~\cite{fallani}. It can also be relevant for electron-hole superfluids in double bilayer graphene devices~\cite{Perali2013} that can have multicomponent effects~\cite{Perali2017}.

The authors acknowledge support from the Brazilian agencies, Conselho Nacional de Ci\^{e}ncia e Tecnologia, CNPq (Grants No. 307552/2012-8, No. 309374/2016-2, and No. 484504/2013-4) and Funada\c{c}\~{a}o de Amparo a Ci\^{e}ncia e Tecnologia do Estado de Pernambuco, FACEPE (APQ-0936-1.05/15). A. P. acknowledges the hospitality and financial support from Universidade Federal de Pernambuco (Grant No. 23076.006072/2015-77) during his visit to Recife in 2015, and also the financial support from the Italian MIUR through the PRIN 2015 program under Contract
No. 2015C5SEJJ001. A. S. and A.V. acknowledge the hospitality of the Departamento de F\'{i}­sica da Universidade Federal de Pernambuco during their temporary stay in Recife in the years 2013 to 2017, and the financial support from the Brazilian agencies, CNPq (Grants No. 452740/2013-4 and 400510/2014-6), and FACEPE (Grant No. ARC 0249-1/13). L. S. acknowledges partial support from FFABR grant of Italian MIUR. A. V. acknowledges support from the Russian Science Foundation under the Project 18-12-00429. The authors thank A. Bianconi, P. Pieri, and
A. A. Varlamov for useful discussions and acknowledge the collaboration within the MultiSuper
International Network (http://www.multisuper.org).

\begin{center}
{\bf SUPPLEMENTAL MATERIALS}
\end{center}

The formalism, used in the manuscript on the screening of fluctuations in two-band superconductors, is outlined. The first section presents details of the relevant mean-field treatment; the second section is focused on the superconducting fluctuations; the third section gives some additional discussions.

\begin{center}
{\bf I - Two-band mean field theory}
\end{center}

We consider a standard microscopic model of a two-band superconductor as described in Refs. \cite{suhl,mos}. The conventional $s$-wave pairing in both bands is controlled by the intra-band interaction strength $g_{\nu \nu }$ ($\nu=1,2$) and the inter-band coupling $g_{12} = g_{21}$ of the Josephson type. We assume one band is deep ($\nu=1$) and the other is shallow $\nu=2$, both bands are two-dimensional with the parabolic dispersion. We consider a system in the clean limit, i.e. without impurity potential. The mean-field Hamiltonian of the model reads as
\begin{align}
{\cal H}= &\sum\limits_{\nu=1,2}\int {\rm d}^2{\bf r}\,
\Bigl[{\hat \psi}^{\dagger}_{\nu\sigma}({\bf r})\,T_\nu({\bf r}) \,{\hat
\psi}_{\nu\sigma}({\bf r})\notag\\
&+ \big({\hat \psi}^{\dagger}_{\nu\uparrow}({\bf
r}){\hat \psi}^{\dagger}_{\nu\downarrow}({\bf r})\, \Delta_{\nu}({\bf r})
+ {\rm h.c.}\big)\Bigr] + (\vec\Delta, \check{g}^{-1}\vec\Delta) , \label{H}
\end{align}
where ${\hat\psi}^{\dagger}_{\nu\sigma}({\bf r})$ and ${\hat\psi}_{\nu\sigma}({\bf r})$ are the field operators for the charge carriers in band $\nu$~($\nu=1,2$), $T_{\nu}({\bf x})$ is the single-particle energy minus the chemical potential. We use the vector notation $\vec\Delta = (\Delta_1,\Delta_2)^T$ with band gaps $\Delta_{\nu =1,2}$ and the scalar product defined as $(\vec{A},\vec{B})=\sum_{\nu}A^*_{\nu}B_{\nu}$. Finally, $\check{g}^{-1}$ is the inverse of the coupling matrix $\check{g}$ with elements $g_{\nu\nu'}$.

The Hamiltonian ${\cal H}$ is to be solved with the self-consistency condition 
\begin{align}
\vec\Delta = \check{g} \vec{R},
\label{self}
\end{align}
where $\vec{R}= (R_1,R_2)^T$ and $R_{\nu}=\langle{\hat \psi}_{\nu\uparrow}({\bf r}){\hat \psi}_{\nu \downarrow}({\bf r})\rangle$, with $\langle\ldots\rangle$ denoting the statistical average. Near the mean-field critical temperature $T_{c0}$ the anomalous Green function $R_{\nu}$ can be approximated as (see, e.g., \cite{vag,orl})
\begin{align}
R_{\nu}[\Delta_{\nu}] \simeq {\cal A}_{\nu}\Delta_{\nu} +\Omega_{\nu}[\Delta_{\nu}],
\label{R}
\end{align}
with
\begin{align}
\Omega_{\nu}[\Delta_{\nu}]= - a_{\nu}\Delta_{\nu} - b_{\nu}\Delta_{\nu}|\Delta_{\nu}|^2 +
{\cal K}_{\nu}\boldsymbol{\nabla}^2\Delta_{\nu},
\label{Omega}
\end{align}
where the coefficients ${\cal A}_{\nu}$, $a_{\nu}$, $b_{\nu}$, and ${\cal K}_{\nu}$ are given by Eq.~(2) in the text. Notice that the zero-field case is considered. Then, the self-consistency condition (\ref{self}) is represented as the matrix gap equation
\begin{align}
\check{L}\vec{\Delta}=\vec{\Omega},
\label{matrixgap}
\end{align}
where $\vec{\Omega}= (\Omega_1,\Omega_2)^T$ and the matrix $\check{L}$ is defined by Eq.~(1) in the manuscript. The solution to Eq.~(\ref{matrixgap}) is the stationary point of the free energy functional given by Eq.~(1) in the manuscript. 

The mean-field transition temperature $T_{c0}$ is obtained from the linearized matrix gap equation $\check{L}\vec{\Delta}=0$ ~\cite{vag, orl} that can be explicitly written as 
\begin{align}
\left(\begin{array}{cc} g_{22} - G {\cal A}_1& -g_{12}\\
				-g_{12}& g_{11} - G {\cal A}_2\end{array}\right) 
				\left(\begin{array}{c} \Delta_1\\
				\Delta_2\end{array}\right) = 0.
\label{lingap}
\end{align}
Notice that $a_{\nu} \to 0$ for $T \to T_{c0}$~[see Eq. (2) of the text] and so the term $a_{\nu} \Delta_{\nu}$ does not contribute to Eq.~(\ref{lingap}). The solution to this equation is represented as 
\begin{align}
\vec{\Delta}=\psi({\bf r})\vec{\eta} ,
\label{lingap1}
\end{align}
where $\vec\eta= (\eta_1,\eta_2)^T$ is an eigenvector of $\check{ L}$ corresponding to its zero eigenstate and $\psi({\bf r})$ is the Landau order parameter and controls the spatial distribution of the both band condensates~\cite{vag,orl}. Notice that the linearized gap equation does not give any information about such a distribution, and one should go beyond the linearized equation to find $\psi$~\cite{vag}. The appearance of the single-component order parameter in the equation for the mean-field critical temperature $T_{c0}$ is connected with the mean-field description of a two-band superconductor by the single-component Ginzburg-Landau (GL) formalism~\cite{vag}. This is in agreement with the Landau theory of phase transitions according to which the number of the order-parameter components is determined by the relevant irreducible group representation rather than by the number of the contributing bands, see discussions in \cite{orl}.  

Introducing the quantity 
\begin{align}
S \equiv (g_{22}-G{\cal A}_1)/g_{12},
\label{S}
\end{align}
the equation $\det\check{L} =0$, which determines $T_{c0}$~ (one chooses the largest $T_{c0}$ of the two solutions), writes as
\begin{align}
1/S= (g_{11}- G {\cal A}_2)/g_{12}.
\label{1/S}
\end{align}
Then, using Eqs.~(\ref{S}) and (\ref{1/S}), $\vec{\eta}$ can be chosen in the form
\begin{align}
\vec{\eta}= \left(\begin{array}{c} S^{-1/2} \\ S^{1/2}\end{array}\right) .
\label{eta}
\end{align}
We remark that the normalization of the eigenvector (\ref{eta}) is arbitrary - it is absorbed in the order parameter $\psi$ in Eq. (\ref{lingap1}). Using Eq.~(2) of the text together with Eqs.~(\ref{S}) and (\ref{1/S}) here, one finds the explicit expressions for $S$ and $T_{c0}$ given by Eqs.~(3) and (4) in the manuscript.

\begin{center}
{\bf II - Fluctuations}
\end{center}

To investigate the superconductive fluctuations in the GL domain, we express superconducting gaps $\Delta_{\nu}$ using $\vec{\eta}$ and $\vec{\xi}$, where $\vec{\xi}$ is any vector orthogonal to $\vec{\eta}$~(the normalization is not important), as
\begin{align}
\vec{\Delta}= \psi \vec{\eta} +\phi \vec{\xi},
\label{exp}
\end{align}
where $\phi=\phi({\bf r})$ is a new spatial mode, additional to $\psi({\bf r})$ introduced in the previous section. Using Eq.~(\ref{exp}), the free energy functional given by Eq.~(1) in the main paper is expressed in terms of $\psi$ and $\phi$, see Eqs.~(5) and (6) in the manuscript. The relevant coefficients in Eq.~(6) of the manuscript are given by
\begin{align}
&a_{\psi} = \sum\limits_{\nu} a_{\nu} |\eta_{\nu}|^2, \; a_{\phi} = (\vec{\xi}, \check{L}\vec{\xi}) +\sum\limits_{\nu} a_{\nu} |\xi_{\nu}|^2,\notag\\
&{\cal K}_{\psi} = \sum\limits_{\nu} {\cal K}_{\nu} |\eta_{\nu}|^2, \;
{\cal K}_{\phi} = \sum\limits_{\nu} {\cal K}_{\nu} |\xi_{\nu}|^2,\notag\\
&b_{\psi} = \sum\limits_{\nu} b_{\nu} |\eta_{\nu}|^4, \; \; b_{\phi} = \sum\limits_{\nu} b_{\nu} |\xi_{\nu}|^4,
\end{align}
where $\eta_{\nu}$ and $\xi_{\nu}$ denote components of $\vec{\eta}$ and $\vec{\xi}$. The key difference between the modes $\psi$ and $\phi$ is that the expression for $a_{\phi}$ contains $(\vec{\xi}, \check{L}\vec{\xi})$, which is nonzero because eigenstates of $\check{L}$ are not degenerate. [The eigenvectors corresponding to the zero eigenvalue form a one-dimensional sub-space, except of the unrealistic case of two equivalent bands with zero inter-band coupling.] Taking into account that $a_{\nu} \to 0$ at $T \to T_{c0}$, one sees that $a_{\psi} \to 0$ but $a_{\phi} \not\to 0$ in this limit. It means that the coherence length associated with the mode $\phi$ is not divergent at $T_{c0}$~(and so is the corresponding contribution to the heat capacity), while the mode $\psi$ is critical and the corresponding length diverges. Thus, investigating the contribution of the superconducting fluctuations near $T_{c0}$, one needs to consider only the critical mode $\psi$. The mode $\phi$ can be safely neglected and one arrives at the single-component GL description of the superconducting transition with the single-component order parameter $\psi$, see the discussion in the previous section after Eq.~(\ref{lingap1}). Using Eq.~(\ref{eta}) of the Supplemental Materials, one can easily get Eq.~(7) of the manuscript. 

To get the Gaussian fluctuation functional, one represents the order parameter $\psi$ as the sum of its ``slow" part $\varphi({\bf r})$~(averaged over fluctuations) and ``fast" contribution $\eta({\bf r})$~(fluctuations).
The standard Gaussian terms in the functional correspond to $|\eta|^2$, $\eta^2$, $\eta^{*2}$, and $|\boldsymbol\nabla\eta|^2$. Then, introducing the real $x_{\bf q}$ and imaginary $y_{\bf q}$ parts of the
Fourier transform of the fluctuation field $\eta_{\bf q}$~($\eta_{\bf q}=x_{\bf q} + \mathbbm{i} y_{\bf q}$), the fluctuation "Hamiltonian" can generally be written as Eq.~(8) in the main paper. The terms involving $x^2_{\bf q}$ and $y^2_{\bf q}$ come from $|\eta|^2$ and $|\boldsymbol\nabla\eta|^2$ whereas $x_{\bf q} x_{-{\bf q}}$ and $y_{\bf q} y_{-{\bf q}}$ result from both $\eta^2$ and $\eta^{*2}$. Performing the standard calculations with the partition function based on Eq.~(8), one finds the fluctuation contribution to the free energy in the form
\begin{align}
F_{\rm fluct}=-\frac{T}{2}\sum\limits_{\bf q}\left[ \ln \frac{\pi T}{A_{\bf q}+B_{\bf q}} +
\ln \frac{\pi T}{A_{\bf q}-B_{\bf q}}\right],
\label{freefluct}
\end{align}
which agrees with the expression in the textbook by Larkin and Varlamov~\cite{lar-var}. In Eq.~(\ref{freefluct}) $A_{\bf q}$ is the momentum-dependent coefficient for the diagonal term $x^2_{\bf q}+y^2_{\bf q}$ in the Gaussian functional whereas $B_{\bf q}$ is the coefficient for the off-diagonal contribution $x_{\bf q} x_{-{\bf q}} - y_{\bf q} y_{-{\bf q}}$, see Eq.~(8) in the manuscript.

To find the explicit expressions for $A_{\bf q}$ and $B_{\bf q}$, one inserts the relation $\psi({\bf r})=\varphi({\bf r}) + \eta({\bf r})$ into the GL equation for $\psi({\bf r})$ and then average the resulting expression over the fluctuations. As explained in the manuscript, when taking into account interactions between the fluctuation fields, the averaging procedure results in Eq.~(10) of the text. Subtracting this equation from the initial GL equation for $\psi$, one finds a rather complicated equation for the fluctuation field $\eta$. However, it is significantly simplified for $T \geq T_c$, where $T_c$ is the fluctuation shifted critical temperature. In this case one finds $A_{\bf q}$ and $B_{\bf q}$ as given by Eq.~(10) of the text.

\begin{center}
{\bf III - Additional Discussions}
\end{center}

{\itshape \underline{Discussion A}}. To estimate an additional shift of the critical temperature due to the Berezinskii-Kosterlitz-Thouless (BKT) fluctuations, we take the fluctuation-averaged order parameter $\varphi({\bf r})$ in the form
\begin{align} 
\varphi({\bf r}) = \varphi_0 \, e^{i\theta({\bf r})}, 
\label{eq:phase}
\end{align}
where $\varphi_0$ is the uniform solution of the fluctuation-averaged GL equation given by Eq.~(9) in the manuscript. The coefficient of the linear term in this equation is expanded in powers of $T-T_c$ and approximated as
\begin{align}
a+2b\langle |\eta|^2 \rangle \approx \widetilde{\alpha} (T-T_c),
\label{eq:app}
\end{align}
with 
\begin{align}
\widetilde{\alpha}= \alpha\,\frac{c^2_0}{2+c_0^2} + {\cal O}(Gi\ln Gi),
\label{eq:alpha}
\end{align}
where $c_0$ is the coefficient appearing in the cut-off $\Lambda_0$, see the manuscript,  It is important to note that $\widetilde{\alpha} > 0$ and, hence, the fluctuation-averaged GL theory is consistent: a nonzero condensate solution appears for $T \leq T_c$. As we consider only the phase fluctuations governed by the Nambu-Goldstone field $\theta({\bf r})$, the corresponding free energy functional is reduced to
\begin{align} 
F = F_0 + \int {\rm d}^2{\bf r}\; \frac{J|\nabla \theta({\bf r})|^2}{2} , 
\label{eq:fluct}
\end{align}
where $F_0$ results from the uniform solution $\varphi_0$ and $J(T)=2 {\cal K}\widetilde{\alpha} (T-T_c)/b$ is the phase stiffness. The compactness of the phase $\theta({\bf r})$ implies that
\begin{align}
\oint_\mathcal{C} {\boldsymbol \nabla} \theta({\bf r}) \cdot 
\mathrm{d}{\bf r} = 2\pi q, 
\end{align}
for any closed contour $\mathcal{C}$. Here $q=0,\pm 1,\pm 2,...$ is the integer number associated with the corresponding quantum vortex (positive $q$) or antivortex (negative $q$). As shown by Kosterlitz and Thouless \cite{Kosterlitz}, in the two-dimensional case ($D=2$), the total number of quantized vortices varies as a function of the temperature: at zero temperature there are no vortices; however, as the temperature increases, vortices start to appear as the vortex-antivortex pairs. The pairs are bound at low temperatures until the Berezinskii-Kosterlitz-Thouless unbinding transition occurs at $T=T_{\rm BKT}$. Above $T_{\rm BKT}$ a proliferation of free vortices and antivortices takes place and the global coherence is destroyed. The BKT temperature $T_{\rm BKT}$ for two-dimensional superconductors can be estimated using the expression~\cite{Kosterlitz}
\begin{align}
T_{\rm BKT} = \frac{\pi}{2} J(\rm T_{BKT}) .
\label{eq:criterion}
\end{align}
From Eq.~(\ref{eq:criterion}) we find 
\begin{align}
{T_c - T_{\rm BKT} \over T_{\rm BKT} } = \frac{4\,\alpha Gi}{\widetilde{\alpha}}={\cal O}(Gi),
\label{eq:shiftbkt}
\end{align}
where $Gi =b/(4\pi\alpha {\cal K})$ is the Ginzburg number associated with the initial GL functional given by Eq.~(6) of the manuscript. Comparing this result with Eq.~(13) of the manuscript, one finds that the additional shift of the critical temperature due to the BKT fluctuations is, of course, important when $Gi$ is large enough but is negligible for the values of the inter-band coupling at which the pseudogap is washed out, i.e., in the limit $Gi \to 0$. Thus, our conclusions are not altered by the inclusion of the fluctuations beyond the Gaussian scheme.

{\itshape \underline{Discussion B}}. As briefly mentioned in the manuscript, we do not claim that the fluctuation screening mechanism, based on the interference of multiple pairing channels with different stiffness, is the only possible scenario to obtain higher critical temperature. Clearly, other effects can give rise to a higher critical temperature. However, whatever those mechanisms might be, the respective transition temperature $T_c$ is calculated within the mean-field approach.

For example, it has been recently demonstrated~\cite{balat} that even within the standard BCS theory the interplay between the interparticle distance and the finite range of the pairing potential gives rise a dome-like shape of the density dependence of $T_c$ with a clearly defined maximum. This can be used to optimize the system for the highest $T_c$. This prediction can be further improved by invoking the Eliashberg approach to take into account the dynamics of the pairing and the Coulomb repulsion.

However, neither of those mean-field mechanisms of reaching a higher $T_c$ takes into account the crucial effect of superconducting fluctuations, which grows in importance with the increasing ratio $T_c/E_F$ ($E_F$ is the Fermi energy). Consequently, one has still to overcome the problem of large fluctuations that always accompany this high mean-field $T_c$. Hence, once the mean field problems toward high $T_c$ are overcome, we remain with the problem of avoiding severe fluctuations you might find in low dimensional systems, in the moderate and strong coupling regime of pairing, and in low density systems. Our work points toward a bright mechanism to solve this ultimate obstacle to reach a very high temperature superconductivity.

{\itshape \underline{Discussion C}}. This work focuses on the regime of the weak-to-moderate pairing coupling. This regime is, however,  relevant even for the room temperature superconductivity. As an example we point to metallo-organic materials (e.g., K-doped paraterphenyl) or the newly discovered class of superhydrites that have dimensionless couplings of the order of $\lambda=0.3$-$0.4$ but very large phononic frequencies, exceeding $100\, {\rm meV} (1160 {\rm K})$. These materials have a multiband electronic structure and reveal Lifshitz transitions \cite{BianJar,Mazziotti2017} and thus are relevant for the analysis of this work. Hence, the coupling of the equation for the critical temperature (given by the Thouless criterion) with the density equation for the chemical potential can be disregarded at this level, being overall the system of our interest at the BCS side of the BCS-BEC crossover. 

In the case of two-band or two-component ultracold atomic fermions, in order to investigate the BCS-BEC crossover in the two-band configuration from the weak-interaction (BCS side) to the strong-interaction (BEC side) regimes, the coupling of the equation for the critical temperature with the density equation for the chemical potential is indeed crucial, differently from the case considered in the manuscript.


\begin{thebibliography}{99}

\bibitem{mp} M.~V.~Milo\v{s}evi\'{c} and A. Perali, Supercond. Sci. and Technol.
{\bf 28}, 060201 (2015).

\bibitem{MgB2a} J. Nagamatsu,  N. Nakagawa, T. Muranaka, Y. Zenitani, and J. Akimitsu,
Nature. {\bf 410}, 63 (2001).

\bibitem{MgB2b} D. C. Larbalestier, L. D. Cooley, M. O. Rikel, A. A. Polyanskii, J. Jiang,
S. Patnaik, X. Y. Cai, D. M. Feldmann, A. Gurevich, A. A. Squitieri, M. T. Naus, C. B. Eom, 
E. E. Hellstrom, R. J. Cava, K. F. Regan, N. Rogado, M. A. Hayward, T. He, J. S. Slusky,
P. Khalifah, K. Inumaru, and M. Haas, Nature {\bf  410}, 186 (2001).

\bibitem{MgB2c} P. C. Canfield and G. W. Crabtree, Physics Today {\bf 56}, 34 (2003).

\bibitem{ironA} Y. Kamihara, H. Hiramatsu, M. Hirano, R. Kawamura, H. Yanagi, T. Kamiya,
H. Hosono, J. Am. Chem. Soc. {\bf 128}, 10012 (2006).

\bibitem{ironB} J. Paglione and R. L. Greene, Nat. Phys. {\bf 6}, 645 (2010).

\bibitem{bian} A. Bianconi, Nat. Phys. {\bf 9}, 536 (2013). 

\bibitem{okaz} K. Okazaki, Y. Ito, Y. Ota, Y. Kotani, T. Shimojima, T. Kiss, S. Watanabe,
C.-T. Chen, S. Niitaka, T. Hanaguri, H. Takagi, A. Chainani, and S. Shin, Sci. Rep. {\bf 4},
4109 (2014).

\bibitem{str1} A. Perali, A. Bianconi, A. Lanzara, N. L. Saini, Solid State Commun. {\bf 100},
181 (1996). 

\bibitem{str2} A. Bianconi, A. Valletta, A. Perali, N. L. Saini, Physica C {\bf 296},
269 (1998).

\bibitem{str3} A. A. Shanenko, M. D. Croitoru, and F. M. Peeters, Phys. Rev. B {\bf 75},
014519 (2007).

\bibitem{str4} N. Pinto, S. J. Rezvani, A. Perali, L. Flammia, M.~V.~Milo\v{s}evi\'{c}‡, M. Fretto,
C. Cassiago, and N. De Leo, Sc. Rep. {\bf 8}, 4710 (2018).
  
\bibitem{eagl} D. M. Eagles, Phys. Rev. {\bf 186}, 456 (1969).

\bibitem{leg} A. J. Leggett, in {\it Modern Trends in the Theory of Condensed
Matter}, edited byA. Pekelski and J. Przystawa (Springer-Verlag, Berlin, 1980), p. 13.

\bibitem{chen} Q. Chen, J. Stajic, S. Tan, and K. Levin, Phys. Rep. {\bf 412}, 1 (2005).

\bibitem{strinati} G. C. Strinati, P. Pieri, G. R{\"o}pke, P. Schuck, and M. Urban,
Phys. Rep. {\bf 738}, 1 (2018).

\bibitem{shan1} A. A. Shanenko, M. D. Croitoru, A. Vagov, and F. M. Peeters, Phys. Rev. B {\bf 82},
104524 (2010). 

\bibitem{shan2} Y. Chen, A. A.  Shanenko, A. Perali, and F. M. Peeters,  J. Phys.: Condens. Matter
{\bf 24}, 185701 (2012).
  
\bibitem{Innocenti2010} D. Innocenti, N. Poccia, A. Ricci, A. Valletta, S. Caprara, A. Perali,
and A. Bianconi, Phys. Rev. B {\bf 82}, 184528 (2010).
  
\bibitem{Mazziotti2017} M. V. Mazziotti, A. Valletta, G. Campi, D. Innocenti, A. Perali,
  and A. Bianconi, Eur. Phys. Lett. {\bf 118}, 37003 (2017). 

\bibitem{guid} A. Guidini and A. Perali, Supercond. Sc. and Technol. {\bf 27}, 124002 (2014). 

\bibitem{bor} S. Borisenko, Nature Mat. {\bf 12}, 600 (2013).

\bibitem{kan} Y. Lubashevsky, E. Lahoud, K. Chashka, D. Podolsky, and A. Kanigel,
Nat. Phys. {\bf 8}, 309 (2012).

\bibitem{kas} S. Kasahara, T. Watashige, T. Hanaguri, Y. Kohsaka, T. Yamashita,
Y. Shimoyama, Y. Mizukami, R. Endo, H. Ikeda, K. Aoyama, T. Terashima, S. Uji,
T. Wolf, H. von L\"ohneysenn, T. Shibauchi, and Y. Matsuda, PNAS {\bf 111}, 16309 (2014).
  

\bibitem{sademelo} C. A. R. S\'{a} de Melo, M. Randeria, and J. R. Engelbrecht,
Phys. Rev. Lett. {\bf 71}, 3202 (1993).  
   
\bibitem{per} A. Perali, P. Pieri, G. C. Strinati, and C. Castellani,  Phys. Rev. B {\bf 66},
024510 (2002).

\bibitem{per-var} A. Perali, C. Castellani, C. Di Castro, M. Grilli, E. Piegari, and A. A. Varlamov,
Phys. Rev. B {\bf 62}, R9295 (2000).

\bibitem{wolf} S. Wolf, A. Vagov, A. A. Shanenko, V. M. Axt, A. Perali, and J. Albino Aguiar,
Phys. Rev. B {\bf 95}, 094521 (2017).
     
\bibitem{suhl} H. Suhl, B. T. Matthias, and L. R. Walker, Phys. Rev. Lett. {\bf 3},
552 (1959).

\bibitem{mos} V. A. Moskalenko, Phys. Met. Metallogr. {\bf 8}, 25 (1959).

\bibitem{geilik} B. T. Geilikman, R. O. Zaitsev, and V. Z. Kresin, Sov. Phys. Solid State {\bf 9}, 642 (1967).

\bibitem{kres} V. Z. Kresin, Journal of Low Temp. Phys. {\bf 11}, 519 (1973).

\bibitem{ashker} I. N. Askerzade, A. Gencer, and N. G\"{u}cl\"{u},
Supercond. Sci. Technol. {\bf 15}, L13 (2002).

\bibitem{zhit} M. E. Zhitomirsky and V.-H. Dao, Phys. Rev. B {\bf 69}, 054508 (2004).

\bibitem{kogan} V. G. Kogan and J. Schmalian, Phys. Rev. B {\bf 83},
054515 (2011).

\bibitem{shan3} A.~A.~Shanenko, M.~V.~Milo\v{s}evi\'{c}, F.~M.~Peeters,
and A.~V.~Vagov, Phys. Rev. Lett. {\bf 106}, 047005 (2011).

\bibitem{vag} A.~Vagov, A.~A.~Shanenko, M.~V.~Milo\v{s}evi\'{c}, V.~M.~Axt,
and F.~M.~Peeters, Phys. Rev. B {\bf 86}, 144514 (2012).

\bibitem{orl} N. V. Orlova, A. A. Shanenko, M. V. Milo\v{s}evi\'{c},
F. M. Peeters, A. V. Vagov, and V. M. Axt, Phys. Rev. B {\bf 87},
134510 (2013).

\bibitem{supp} See also the Supplemental Materials.

\bibitem{popov} V. N. Popov, {\it Functional Integrals in Quantum Field Theory
and Statistical Physics}, (Springer Netherlands, 1983). 

\bibitem{lar-var} A. Larkin and A. Varlamov, {\it Theory of Fluctuations
in Superconductors}, (Oxford Univ. Press, Oxford, 2005).

\bibitem{zaremba} A. Griffin, T. Nikuni, and E. Zaremba, {\it Bose-Condensed Gases at
Finite Temperatures} (Cambridge Univ. Press, Cambridge, 2009).

\bibitem{NKcrit} D. R. Nelson and J. M. Kosterlitz, Phys. Rev. Lett. {\bf 39}, 1201 (1977).

\bibitem{vag1} A. Vagov, A. A. Shanenko, M. V. Milo\v{s}evi\'{c}, V. M. Axt,
V. M. Vinokur, J. Albino Aguiar, and F. M. Peeters, Phys. Rev. B {\bf 93}, 174503
(2016).

\bibitem{kett} J. B. Ketterson and S. N. Song, {\itshape Superconductivity}
(Univ. Press, Cambridge, 1999).

\bibitem{Hanag} T. Hanaguri, S. Kasahara, J. B\"{o}ker, I. Eremin, T. Shibauchi, Y. Matsuda,
arXiv: 1901.09141, Jan 2019, ``Quantum vortex core and missing pseudogap in the multi-band BCS-BEC-crossover superconductor ${\rm FeSe}$".

\bibitem{Balatsky} E. Langmann, C. Triola, and A. V. Balatsky, arXiv:1810.03349, Oct 2018, ``Ubiquity of superconductivity domes in BCS theory with finite-range potentials".

\bibitem{iskin} M. Iskin and C. A. R. S\'{a} de Melo, Phys. Rev. B {\bf 74}, 144517 (2006).
   
\bibitem{tempere} S. N. Klimin, J. Tempere, G. Lombardi, J. T. Devreese,
  Eur. Phys. J. B {\bf 88}, 122 (2015). 
  
\bibitem{fallani} G. Pagano, M. Mancini, G. Cappellini, L. Livi, C. Sias, J. Catani, M. Inguscio,
and L. Fallani, Phys. Rev. Lett. {\bf 115}, 265301 (2015).

\bibitem{Perali2013} A. Perali, D. Neilson, and A.R. Hamilton, Phys. Rev. Lett. {\bf 110}, 146803 (2013).

\bibitem{Perali2017} S. Conti, A. Perali, F. M. Peeters, and D. Neilson,
Phys. Rev. Lett. {\bf 119}, 257002 (2017).

\bibitem{Kosterlitz} J. M. Kosterlitz and D.J. Thouless, J. Phys. C: Solid State Phys. {\bf 6}, 1181 (1973).
  
\bibitem{balat} E. Langmann, C. Triola, and A. V. Balatsky, arXiv:1810.03349, Oct 2018, ``Ubiquity of superconductivity domes in BCS theory with finite-range potentials".

\bibitem{BianJar} A. Bianconi and T. Jarlborg, Eur. Phys. Lett.  {\bf 112}, 37001 (2015).

\end{thebibliography}
\end{document}